\documentclass[graybox,envcountchap,sectrefs]{svmono}

\usepackage[ruled,vlined,linesnumbered,noresetcount]{algorithm2e} %noend

\usepackage{type1cm}         

\usepackage{makeidx}         % allows index generation
\usepackage{graphicx}        % standard LaTeX graphics tool
\usepackage{multicol}        % used for the two-column index
\usepackage[bottom]{footmisc}% places footnotes at page bottom

\usepackage{newtxtext}       % 
\usepackage[varvw]{newtxmath}       % selects Times Roman as basic font

\usepackage{multirow} %for multi-row table

\usepackage{subcaption}

\usepackage{hyperref}
\hypersetup{
    colorlinks=true,
    linkcolor= cyan,
    filecolor=blue,
    urlcolor=blue,
    citecolor=blue
}
\usepackage[numbers]{natbib}
\usepackage{chapterbib}
\usepackage[acronym]{glossaries}

\theoremstyle{plain}
\newtheorem{Example}{Example}[section] %with section, its index will come along with the section index. 

\makeglossaries
\newacronym{cps}{CPS}{Cyber-Physical System}
\newacronym{hcps}{HCPS}{Human-Cyber-Physical System}
\newacronym{hmi}{HMI}{Human Machine Interface}
\newacronym{apt}{APT}{Advanced Persistent Threat}
\newacronym{dos}{DoS}{Denial-of-Service}
\newacronym{idos}{IDoS}{Informational Denial-of-Service}
\newacronym{ddos}{DDoS}{Distributed Denial-of-Service}
\newacronym{hra}{HRA}{Human Reliability Analysis}
\newacronym{cdf}{CDF}{Cumulative Distribution Function}
\newacronym{ids}{IDS}{Intrusion Detection System}

\newacronym[% options to override defaults
  longplural={Areas of Interest}
]
{aoi}{AoI}{Area of Interest}

\newacronym{bo}{BO}{Bayesian Optimization}

\newacronym{ltm}{LTM}{Long-Term Memory}

\newacronym{ai}{AI}{Artificial Intelligence}
\newacronym{rl}{RL}{Reinforcement Learning}
\newacronym{mdp}{MDP}{Markov Decision Process}

\newacronym{se}{SE}{Social Engineering}
\newacronym{soc}{SOC}{Security Operation Center}

\newacronym{sa}{SA}{Situation Awareness}
\newacronym{ttp}{TTP}{Tactics, Techniques, and Procedures}
\newacronym{cpt}{CPT}{Cumulative Prospect Theory}
\newacronym{eut}{EUT}{Expected Utility Theory}

\newacronym{ne}{NE}{Nash Equilibrium}
\newacronym{smdp}{SMDP}{Semi-Markov Decision Process}
\newacronym{smg}{SMG}{Semi-Markov Game}

\newacronym{tom}{ToM}{Theory of Mind}

\usepackage{imakeidx}

\makeindex             % used for the subject index
\begin{document}

\author{Linan Huang and Quanyan Zhu}
\title{Cognitive Security}
\subtitle{-- A System-Scientific Approach --}
%\maketitle

%\frontmatter%%%%%%%%%%%%%%%%%%%%%%%%%%%%%%%%%%%%%%%%%%%%%%%%%%%%%%

%\include{preface}

% \include{author/dedication}
% \include{author/foreword}
% \include{author/preface}
% \include{author/acknowledgement}

\setcounter{secnumdepth}{3}
\setcounter{tocdepth}{3}
%\tableofcontents
% \listoftables\addcontentsline{toc}{section}{\listtablename}

%\include{acronym}
%\printglossary[type=\acronymtype]
%\addcontentsline{toc}{section}{List of Acronyms}

%\newpage

%\mainmatter%%%%%%%%%%%%%%%%%%%%%%%%%%%%%%%%%%%%%%%%%%%%%%%%%%%%%%%
%\include{part}

%%%%%%%%%%%%%%%%%%%%% chapter.tex %%%%%%%%%%%%%%%%%%%%%%%%%%%%%%%%%
%
% sample chapter
%
% Use this file as a template for your own input.
%
%%%%%%%%%%%%%%%%%%%%%%%% Springer-Verlag %%%%%%%%%%%%%%%%%%%%%%%%%%
%\motto{Use the template \emph{chapter.tex} to style the various elements of your chapter content.}
\chapter{
\huge{An Introduction of System-Scientific Approaches to Cognitive Security}
\\
\quad \\
\Large{Linan Huang and Quanyan Zhu}}
\label{chap:intro} % Always give a unique label
\chaptermark{An Introduction of System-Scientific Approaches to Cognitive Security}

% to alter or adjust the chapter heading in the running head

% \abstract*{Each chapter should be preceded by an abstract (no more than 200 words) that summarizes the content. The abstract will appear \textit{online} at \url{www.SpringerLink.com} and be available with unrestricted access. This allows unregistered users to read the abstract as a teaser for the complete chapter.
% Please use the 'starred' version of the new \texttt{abstract} command for typesetting the text of the online abstracts (cf. source file of this chapter template \texttt{abstract}) and include them with the source files of your manuscript. Use the plain \texttt{abstract} command if the abstract is also to appear in the printed version of the book.}

%\title{An Introduction of System-Scientific Approaches to Cognitive Security}
%\author{Linan Huang and Quanyan Zhu}

\begin{figure}[tbh]
\sidecaption[t]
\includegraphics[width=1 \textwidth]{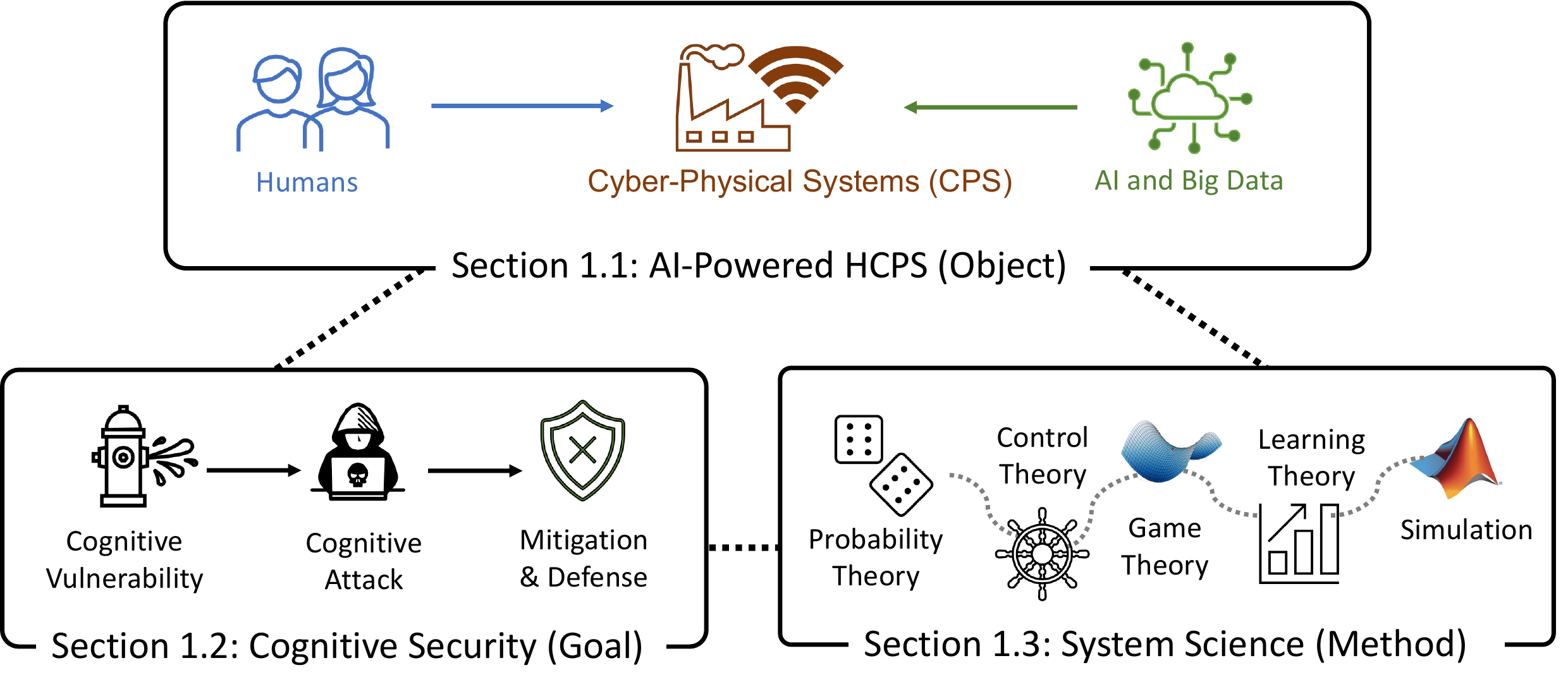}
\caption{
The structure of Chapter \ref{chap:intro} concerning  \acrshort{ai}-powered HCPS in Section \ref{sec: AI-Powered Human-Cyber-Physical Systems}, cognitive security in Section \ref{sec:Cognitive Security in HCPS}, and system science in Section \ref{sec:System-Scientific Perspectives for Cognitive Security} as the object, goal, and method of this book, respectively. 
}
\label{fig:chap1structure}      
\end{figure}

%<200 words
\abstract{
Human cognitive capacities and the needs of human-centric solutions for ``\textit{Industry 5.0}'' make humans an indispensable component in \acrfullpl{cps}, referred to as \acrfullpl{hcps}, where  \acrshort{ai}-powered technologies are incorporated to assist and augment humans. 
The close integration between humans and technologies in Section \ref{sec: AI-Powered Human-Cyber-Physical Systems} and cognitive attacks in Section \ref{subsec:Cognitive Attack} poses emerging security challenges, where attacks can exploit vulnerabilities of human cognitive processes, affect their behaviors, and ultimately damage the \acrshort{hcps}. 
\newline\indent
Defending \acrshortpl{hcps} against cognitive attacks requires a new security paradigm, which we refer to as ``\textit{cognitive security}'' in Section \ref{subsec:Cognitive Security}.  
The vulnerabilities of human cognitive systems and the associated methods of exploitation distinguish cognitive security from ``\textit{cognitive reliability}'' and give rise to a distinctive \textit{CIA triad}, as shown in Sections \ref{subsubsec:Cognitive Reliability vs. Cognitive Security} and \ref{subsubsec:CIAtriad}, respectively.  Section \ref{subsubsec:Cognitive and Technical Defenses} introduces cognitive and technical defense methods that deter the kill chain of cognitive attacks and achieve cognitive security. 
System scientific perspectives in Section \ref{sec:System-Scientific Perspectives for Cognitive Security} offer a promising direction to address the new challenges of cognitive security by developing \textit{quantitative, modular, multi-scale, and transferable} solutions. 
Fig. \ref{fig:chap1structure}  illustrates the structure of Chapter \ref{chap:intro}.
}
 % list of 6-10 keywords for each chapter.

\section{AI-Powered Human-Cyber-Physical Systems}
\label{sec: AI-Powered Human-Cyber-Physical Systems}
\begin{figure}[tbh]
\sidecaption[t]
\includegraphics[width=1 \textwidth]{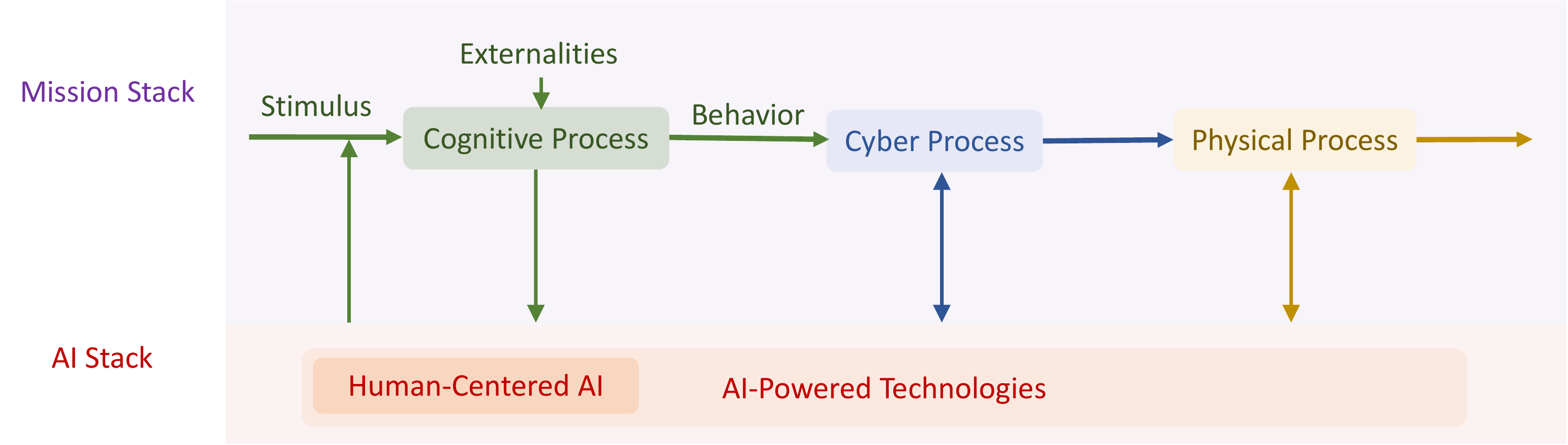}
\caption{
The overview diagram of human roles and  \acrshort{ai}-powered technologies in \acrfullpl{cps}. 
The mission stack at the top illustrates the task-driven flow chart that consists of human, cyber, and physical processes in green, blue, and yellow, respectively. 
%Consists of human, cyber, and physical process in green, blue, and yellow, respectively, the mission stack illustrates the flow chart to accomplish a mission. 
The  \acrshort{ai} stack at the bottom illustrates the  \acrshort{ai}-powered technologies that interact with the human-cyber-physical process to assist with mission completion. 
The \acrshort{ai}-powered technologies are designed to be scalable and transferable. In addition, they need to be  customized, explainable, ethical, respectful of privacy, and trustworthy to achieve human-centric objectives. 
%additional requirement of explainability and ethically aligned . 
}
\label{fig:TwoTrends}       % Give a unique label
\end{figure}

\acrfullpl{cps} are ``smart systems that include engineered interacting networks of physical and computational components'', as defined by the National Institute of Standards and Technology (NIST) in $2017$ \cite{griffor2017framework}. 
Despite the increasing automation and intelligence in \acrshortpl{cps}, humans play indispensable roles in accomplishing \acrshort{cps} tasks, as illustrated by the mission stack in Fig. \ref{fig:TwoTrends}. 
The rapid development of \acrfullpl{ai} and big data has facilitated a close integration of  \acrshort{ai}-powered technologies along with the mission completion, as illustrated by the  \acrshort{ai} stack in Fig. \ref{fig:TwoTrends}.  
We elaborate on the human-involved mission stack and the  \acrshort{ai} stack in Sections \ref{subsec:human roles in CPS} and \ref{subsec:mission stack and  AI stack}, respectively.  

\subsection{Human Roles in Mission Stack}
\label{subsec:human roles in CPS}
Humans \textit{cannot} and \textit{should not} be superseded in \acrshortpl{cps}. 
On the one hand, due to their distinguished cognitive power and analytical capabilities (e.g., logical reasoning, symbolic abstraction, knowledge transfer, understanding others' intent and emotions, etc.), humans have been playing irreplaceable roles, including determining demands, designing mechanisms, and responding to incidents. 
%vision or unavoidable progress
On the other hand, the inevitable progress of ``Industry 5.0'' \cite{industry5} has reemphasized human-centric solutions to create personalized products, services, and experiences. 
The increased level of automation aims to \textit{support} rather than \textit{supersede} humans.  

The mission stack in Fig. \ref{fig:TwoTrends} summarizes the system-level integration of human touch with cyber and physical processes, referred to as \acrfullpl{hcps}, for mission completion. 
After a human participant receives a stimulus, he processes it and outputs responsive behaviors that influence the cyber and physical processes. 
The cognitive process can be affected by externalities related to individual factors (e.g., personality, awareness, and expertise), environmental factors (e.g., workload and stress), and social factors (e.g., peer pressure and culture). 

Human participants have different roles (e.g., users, operators, security analysts, and administrators) in mission completion, and the associated cognitive process can take different forms. Some examples are provided as follows. 
\begin{description}[Example 1]
    \item[Example 1] {End users, including employees and contractors, use the computing facilities of the providers to maintain the corporation's normal operations. 
    For example, the stimulus for employees could be working emails. Their cognitive processes affect the  accuracy and timeliness of the phishing\footnote{Phishing is coined as a combination of the words ``password'' and ``fishing'' to describe the practice of tricking Internet users into revealing sensitive data, including passwords.} recognition, which results in either secure behaviors or falling victim to phishing. }
    \item[Example 2] {Security analysts investigate alerts in real time for alert triage and response. 
    The alerts, the triage process, and the response are the stimulus, the cognitive process, and the behaviors, respectively, in this alert management scenario.}
    %\item Doctor remote surgery. 
\end{description}

%More details of cognitive capabilities and human roles in mission stack of \acrshortpl{hcps} can be found in Chapter \ref{chap:human roles}. 

\subsection{Incorporating  AI Stack into Mission Stack}
\label{subsec:mission stack and  AI stack}

The advances in \acrshort{ai} have accelerated the automation and smartification process in cyber and physical layers, as shown by the blue and yellow double-headed arrows in Fig. \ref{fig:TwoTrends}. 
\acrshort{ai} has been widely applied to sensing, control, communication, and data processing in a variety of applications such as biomedical monitoring, robotics systems, and digital twins \cite{Song2022,salau2022recent,groshev2021toward,doghri2022cyber}. 
In these \acrshort{cps} applications, \acrshort{ai}  not only serves as a technology for reasoning, planning, learning, and processing, but also enables the manipulation of physical objects. 
For example, autonomous driving cars adopt \acrshort{ai} to sense the environment (e.g., road condition, weather, and the movement of pedestrians and other cars) and determine the optimal driving setting (e.g., speed, brake, and steer). 
 \acrshort{ai}-powered technologies in these \acrshort{cps} applications should be scalable \cite{Barmer2021} and transferable \cite{pan2009survey}. 

%first a general description of human-center  \acrshort{ai}'s function and benefit to assist human
Compared to \textit{technology-based}  \acrshort{ai} design that enables humans to adapt to the technical system, \textit{human-centered}  \acrshort{ai} aims to design systems that are aware of human cognitive processes to augment human perception, cognition, and decision-making capabilities in the rapidly evolving, complex, and uncertain \acrshort{cps} environment. 
Besides the desirable features of scalability and transferability for  \acrshort{ai}-powered technologies, we further require human-centered  \acrshort{ai} to be customized, explainable, ethical, respectful of privacy, and trustworthy. 
Such requirements are based on a thorough understanding of the human cognitive process and fulfilled by designing proper human-assistive technologies (e.g., \acrfullpl{hmi}), as shown by the down-side and up-side green arrows in Fig. \ref{fig:TwoTrends}, respectively. 

The design of human-centered  \acrshort{ai} should also be adaptive to \acrshort{cps} applications of different functions, features, requirements, and constraints. 
For example, it is essential for a driving-assistive system to be explainable to facilitate trust and minimize a driver's decision time to take proper action \cite{chaczko2020exploration}. 
Meanwhile, an  \acrshort{ai}-enabled \acrshort{hmi} in the control room of a power grid should feature the following functions \cite{marottowards}: 
\begin{itemize}
    %\item Show contextually relevant information at the right time due to the time-constrained environment
    \item Supporting efficient invocation and dismissal because the number of actions an operator can do in a time window is limited, especially in a time-constrained environment.  
    \item Taking into account and learn from user behavior and feedback because grid operators are well-trained experts, capable of evaluating the assistant’s answers and providing feedback. 
    \item Conveying the consequences of user actions because operators may not be able to estimate the risk correctly and timely. 
\end{itemize}
%A sketch of AI and learning methods is presented in Section \ref{sec:learning theory}, and two examples of applying human-centered \acrshort{ai} into \acrshort{hcps}. 

\section{Cognitive Security in HCPS}
\label{sec:Cognitive Security in HCPS}

%In Section \ref{sec: \acrshort{ai}-Powered Human-Cyber-Physical Systems}, we introduce the mission stack and the  \acrshort{ai} stack to fulfil CPS missions. 
\acrshortpl{cps} have been under threat of various attacks since their emergence.  
Established on the mission stack and the  \acrshort{ai} stack illustrated in Fig. \ref{fig:TwoTrends}, we incorporate the attack stack and the defense stack in Section \ref{subsec:Attack and Defense Stack} to form the four-stack dissection of \acrshort{hcps} security, as illustrated in Fig. \ref{fig:4stack}. 
After the above panorama of \acrshort{hcps} security, we zoom into the focus of his book, i.e., cognitive process, cognitive vulnerability, cognitive attack, and cognitive security, in Sections \ref{subsec:Cognitive Process}, \ref{subsec:Cognitive Vulnerability}, \ref{subsec:Cognitive Attack}, and \ref{subsec:Cognitive Security}, respectively.

\subsection{Attack and Defense Stack}
\label{subsec:Attack and Defense Stack}

\begin{figure}[tbh]
\sidecaption[t]
\includegraphics[width=1 \textwidth]{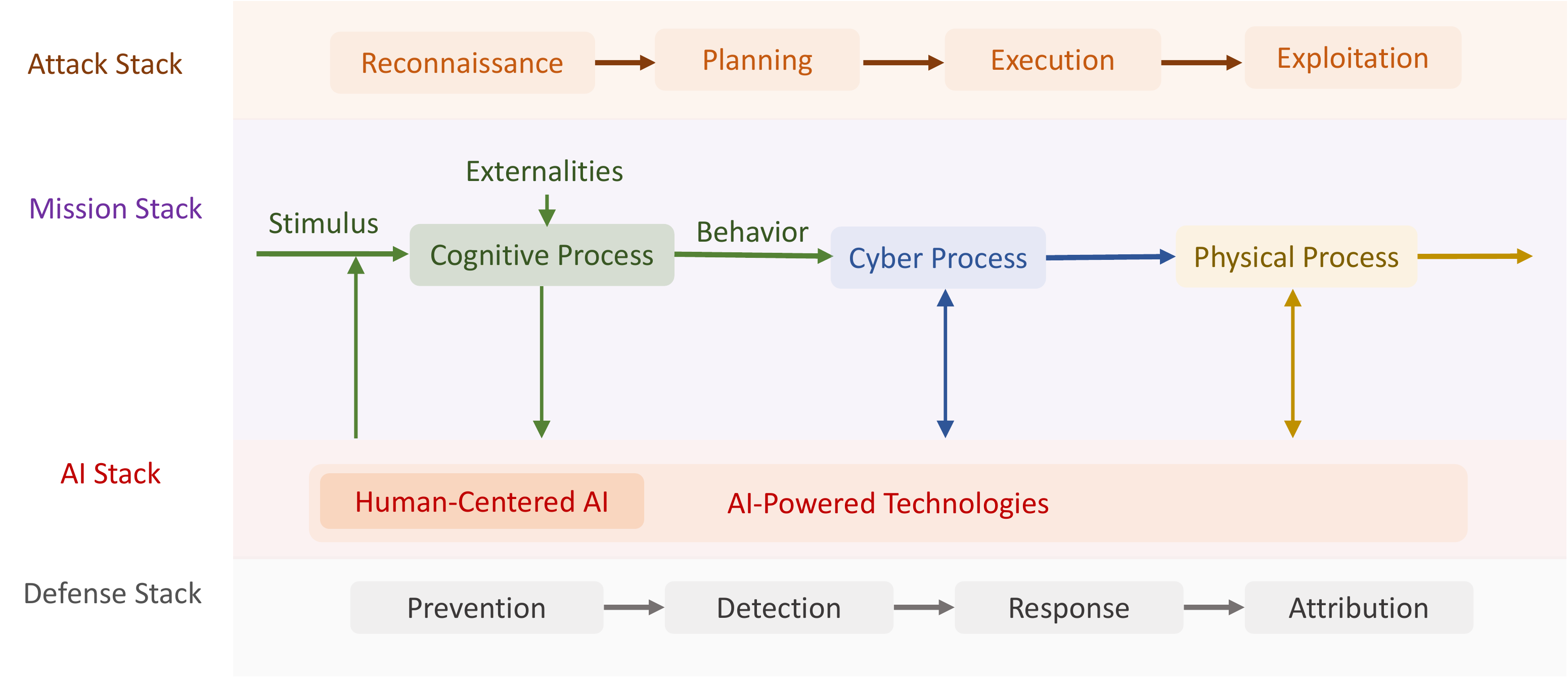}
\caption{
The vertical dissection of \acrshort{hcps} security consists of attack and defense stacks over the mission and  \acrshort{ai} stacks. 
The four orange boxes in the attack stack illustrate the typical life cycle of an attack. 
First, attackers tailor their attack strategies and select attack tools in the planning stage based on the reconnaissance results (e.g., exploitable human and technical vulnerabilities). 
Then, an attack is launched and inflicts damage in the execution and exploitation stages, respectively. 
As illustrated by the four gray boxes in the defense stack, a typical defense consists of four stages: prevention, detection, response, and attribution. 
}
\label{fig:4stack}       % Give a unique label
\end{figure}

Fig. \ref{fig:4stack} illustrates the vertical dissection of \acrshort{hcps} security, including the attack stack and the defense stack in orange and gray, respectively. 
In the past decade, attacks have evolved to be targeted, intelligent, multi-staged, and multi-phased. 
%which is referred to as \acrfullpl{apt}. 
The typical life cycle of these attacks consists of the following four stages. 
In the reconnaissance stage, attackers gather intelligence to determine the attack goal and identify vulnerabilities in the target network. In the planning stage, attackers tailor their strategies to the selected vulnerabilities and choose attack tools. In the execution stage, the attackers deploy the malware to gain an initial foothold, expand access, gain credentials, and escalate privilege. Finally, the attack in the exploitation stage exfiltrates confidential data, interrupts cyber services, or inflicts physical damage.

The four gray boxes in the defense stack illustrate four defensive courses of action. The prevention stage includes the precautionary and proactive defense methods used in advance of attacks. Intrusion prevention techniques, including firewalls and demilitarized zones (DMZ), may be ineffective, especially for advanced attacks such as \acrfullpl{apt}. Therefore, intrusion detection and response are necessary to protect against them. The attribution stage includes post-event analysis, threat intelligence acquisition, and an accountability system.

\subsection{Feedforward and Feedback Cognitive Processes}
\label{subsec:Cognitive Process}

\begin{figure}[tbh]
\sidecaption[t]
\includegraphics[width=1 \textwidth]{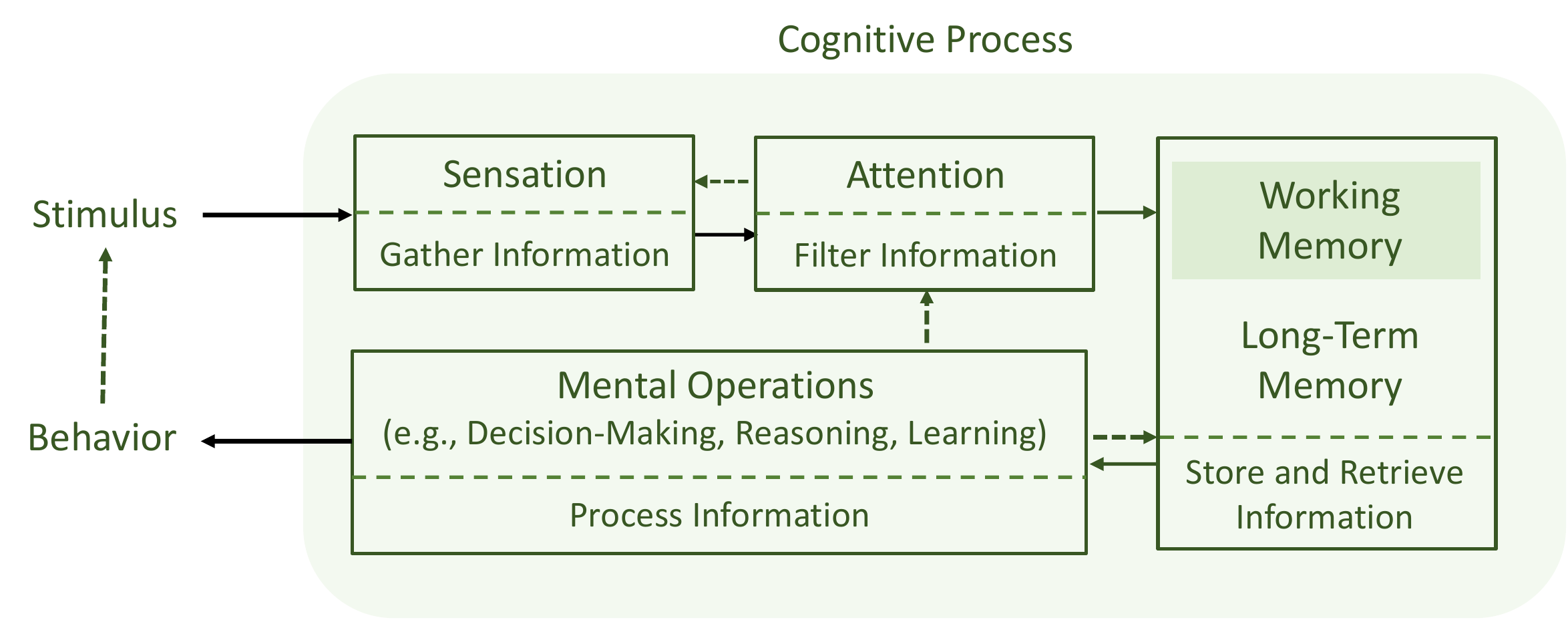}
\caption{
A zoomed-in version of the cognitive process that includes sensation, attention, memory, and mental operations.  
The feedforward path in solid arrows illustrates the information flow of gathering, filtering, storing, retrieving, and processing. 
In the feedback loops, behaviors alter future stimuli, mental operations affect \acrshort{ltm} and attention, and selective attention determines perception, respectively, as illustrated by the dashed arrows. 
}
\label{fig:CogProZoomin}      
\end{figure}

Fig. \ref{fig:CogProZoomin} provides a zoomed-in version of the dynamic cognitive process that consists of feedforward and feedback processes, represented by solid and dashed arrows, respectively. 
We first illustrate the feedforward cognitive process, where the stimulus is the input and the behavior is the output. 
First, perception \cite{bruce2003visual} decodes and gathers the sensory information from external stimuli. 
Due to the limited cognitive capacity, attention is used to filter information and store essential or urgent items in working memory. 
Then, humans process the information for mental operations, including decision-making, reasoning, and learning, which lead to behaviors. 
Based on the tasks and scenarios, human mental operations may need to retrieve past experiences that are stored in the \acrfull{ltm}
The cognitive process also includes several feedback loops listed below:  
\begin{itemize}
    \item Behaviors have an impact on the \acrshort{cps} and change the subsequent stimulus. 
    \item Mental operations learn new knowledge and store experience in the \acrshort{ltm}. 
    \item Besides passively filtering the collected information in the feedforward process, attention also actively affects perception by directing our awareness to relevant stimuli while suppressing distracting information. Such \textit{selective attention} \cite{johnston1986selective} has been demonstrated in phenomena such as the \textit{cocktail party effect} \cite{arons1992review}. 
    \item \textit{Selective attention} is a result of the mental operations and is usually goal-driven and endogenous (referred to as the \textit{top-down attention}), compared to the \textit{bottom-up attention} that is endogenously driven by the stimuli.  
\end{itemize}

\subsection{Exploitable Cognitive Vulnerability}
\label{subsec:Cognitive Vulnerability}

Each component of the cognitive process in Fig. \ref{fig:CogProZoomin} possesses distinct vulnerabilities exploitable by attackers. 
\cite{cybenko2002cognitive}
We discuss vulnerabilities concerning perception, attention, memory, and mental operations in Sections \ref{subsubsec:v_perception}, \ref{subsubsec:v_attention}, \ref{subsubsec:v_memory}, \ref{subsubsec:v_mental}, respectively. 
%We refer the readers to Chapter \ref{chap:SS model of cognition in CPS} for a detailed review of system-scientific approaches to analyze, exploit, and mitigate these cognitive vulnerabilities. 

\subsubsection{Vulnerability of Perception}
\label{subsubsec:v_perception}

Perception is a complex process that involves visual, auditory, somatosensory, olfactory, gustatory, and vestibular systems. 
%Due to the hysteresis effect of human perception, the 
%spanning from audio-visual interpretation to features integration and pattern recognition \cite{burda2021dissecting}.  
These systems follow patterns that can be exploited by attackers. 
For example, knowing the perception limits (e.g., it takes approximately $0.3$ and $1$ seconds to see and hear a signal, respectively) and contributing factors (e.g., how light, color, and noise affect human perception), attackers can create environments where humans are prone to perception errors. 
Human perception systems can sometimes undergo distortions of the senses (e.g., visual and auditory illusions) and produce false sensory information. 
By understanding the contributing factors and causes of these sensory illusions, attackers can craft phishing websites and emails with fewer identifiable phishing indicators. 

Besides exploitable patterns and illusions, human perception is also susceptible to manipulation. 
%The manipulation of perception can be either covert (e.g., the subtle manipulation of perceptions and the blatant use of misleading information) or overt (e.g., defacing or spoofing legitimate forms of communication to influence the user), depending on whether the attacker tries to disguise the attack \cite{cybenko2002cognitive}. 
Attacks have adopted many psychological techniques, including priming \cite{molden2014understanding}, to manipulate perception in \acrshortpl{hcps}.  
%referred to as the digital nudging. %this is a positive word, use nudging for defense purpose
Priming is a well-known phenomenon in psychology wherein the presence of a stimulus impacts how later stimuli are processed or interpreted (e.g., humans recognize the word ``food'' more quickly after the presentation of the word ``kitchen'' than the word ``office'')  \cite{lindsay2020attention}. 
The majority of priming is positive and increases the sensitivity of the related stimulus, while negative priming \cite{tipper1985negative} slows down the processing speed. 
Both \textit{positive} and \textit{negative} priming can be weaponized for perception hacking. For example, attacks may use positive priming to emphasize certain ideas in phishing or use negative priming to deemphasize the phishing indicators. 
Based on whether the stimulus is consciously perceptible or not, priming is classified as \textit{supraliminal} and \textit{subliminal}, respectively \cite{elgendi2018subliminal}.  
Due to its stealthiness, subliminal priming can be a primary candidate for perception hacking. 
As shown in the experiment results \cite{huang2020detecting}, by subtly presenting words or images that are physically or semantically similar to the judgments preferred by the attackers, attackers can influence the accuracy or false alarm rate of the inspectors.

\subsubsection{Vulnerability of Attention}
\label{subsubsec:v_attention}

Attention can be described as an overall level of alertness or ability to engage with surroundings \cite{lindsay2020attention}. 
Humans rely on their attention mechanisms to control their limited computational resources in the brain for the maximum information acquisition. 
Despite the remarkable success of human attention, it suffers from reduced performance under multitasking, long duration, stress, fatigue, and heavy cognitive load. 
%\citet{mack1998inattentional} coins the term `inattentional blindness' to describe the phenomena of individuals failing to  unexpected but often salient stimulus, purely as a result of a lack of attention rather than any vision defects or deficits. 
Moreover, as the result of the selectiveness of our attention to prevent us from getting lost in irrelevant information, we can go through failures of selection in space and time. 
On the one hand, failures of selection in space has been demonstrated in experiments of \textit{change blindness} \cite{simons2005change} and \textit{change deafness} \cite{vitevitch2003change}, where observer does not notice a change in a visual and auditory stimulus, respectively. 
On the other hand, experiments of \textit{attentional blink} \cite{shapiro1997attentional} and \textit{repetition blindness} \cite{kanwisher1987repetition} have shown that failures can occur along with time; i.e., when new information (even of small amount) continues to arrive, processing it leads to the miss of other information. 

Attacks can exploit these \textit{spatial} and \textit{temporal} attentional vulnerabilities either reactively or proactively. 
\textit{Reactive} attention attacks exploit inattention to evade detection and do not attempt to change human attention patterns. For example, many \acrfull{se} and phishing attacks result from a lack of attention. We provide a defense framework against reactive attention attacks in \cite{huang2022advert}.
%(e.g., careless users fail to notice the phishing indicators, including spelling errors and grammar mistakes) and inattentional blindness (e.g., users focusing on the main content fail to perceive unloaded logos in a phishing email \cite{Benishti_2020}).
In contrast, \textit{proactive} attention attacks aim to strategically influence human attention patterns. 
For example, an attacker can generate a large volume of feints and hide real attacks among them to overload human operators, delay their responses, and reduce the accuracy of their judgements. 
We refer to this new form of attacks as the \acrfull{idos} attacks and present a formal description of \acrshort{idos} attacks, their impacts, and associated defense methods in  \cite{huang2022radams}.

\subsubsection{Vulnerability of Memory}
\label{subsubsec:v_memory}

Relying on networks of neurons in the brain, human memory suffers from restricted capacity, limited speed of information storage and retrieval, forgetting, and memory errors. 
While digital storage devices share the first two memory vulnerabilities, the latter two are unique to human memory. 
According to \citet{schacter2002seven}, the latter two belong to the \textit{sins of omissio} and \textit{commission}, respectively. 

Forgetting is the spontaneous or gradual loss of information already stored in an individual's short- or long-term memory.  
Unlike a digital storage device, humans cannot `forget on demand'; i.e., items will linger in memory even then they are no longer needed \cite{sasse2001transforming}. 

Memory errors refer to the wrong recall of information. This can include remembering things that have not happened, giving the wrong source for a memory, or making up things that did not happen. %or occur in a different way from the memory. 
Memory errors are caused in part by the structure of neuron networks as well as a feature of human memorization. 
As shown in the Deese–Roediger–McDermott paradigm \cite{deese1959prediction}, humans incorrectly recall an absent word as it is related to a list of words that belong to a common theme. 
Many factors (e.g., the degree of attention, motivation, emotional state, and environment where memorization takes place) can affect human memory. 
For example, the \textit{emotional enhancement of memory} \cite{hamann2001cognitive} has demonstrated that emotional stimuli are more easily remembered than neutral stimuli. 

Human memory vulnerabilities directly lead to security risks. For example, humans use simple and meaningful passwords, reuse the same password over different sites, and even write down the passwords to remember them, which makes the passwords insecure. 
As will be introduced in Definition \ref{def:Cognitive Reliability}, cognitive reliability methods, including Single Sign-On (SSO)\footnote{Single SSO allows a user to log in several related systems with a single ID and password, it reduces the total number of passwords to remember.}  \cite{radha2012survey}, cognitive passwords \cite{zviran1990cognitive}, and graphical passwords \cite{biddle2012graphical}, are introduced to mitigate memory vulnerability concerning passwords. 

Attackers can also actively exploit those memory vulnerabilities and manipulate the above factors to create attack vectors. 
For example, attackers can reduce the attack frequency to exploit the forgetting vulnerability. 
As demonstrated in \cite{sawyer2018hacking} and \cite{kaivanto2014effect}, lower frequency and likelihood of phishing events increase victim’s susceptibility to phishing cyberattacks. 
%Insight 11 The success of social engineering cyberattacks is inversely related to their prevalence. 
Due to the \textit{suggestibility} of human memory (i.e., humans are inclined to accept and act on the suggestions of others), attackers can design phishing emails to trigger memory errors and inject false memories by designing misleading hints. 
Moreover, they can use emotional language to enhance the operator's false memory and facilitate trust.

\subsubsection{Vulnerability of Mental Operations}
\label{subsubsec:v_mental}

Mental operation vulnerabilities primarily refer to a variety of cognitive biases and exploitable traits. 
In the history of human development, we have developed cognitive shortcuts and biases for rapid, although less accurate or reasonable, responses to survive in highly dynamic and uncertain environments. 
However, those cognitive biases expand the \textit{attack surface} and make humans susceptible to \acrshort{se} attacks. 
We list some of the cognitive biases and the potential adversarial exploitation \cite{Rohleder_2019} as follows. 
\begin{itemize}
    \item \textbf{Anchoring}: An individual's decisions are influenced by a particular reference point (i.e., an ``anchor''). Pretexting \acrshort{se}  can create a situation and initial context to increase the attack’s apparent legitimacy and the likelihood of success. Anchoring bias keeps people from questioning the initial impression and accepting the scam. 
    \item \textbf{Framing}: An individual can draw different conclusions from the same information, depending on how that information is presented (e.g., as a loss or a gain). Attackers can utilize the framing effect to craft the content of phishing emails and distort human risk perception. 
    \item \textbf{Optimism bias}: People tend to have unrealistic optimism; e.g.,  overestimating the likelihood of positive events and underestimating the likelihood of negative events. 
    Users frequently believe others are more susceptible than they are \cite{cox2020stuck}. 
    Since they think they are immune to attacks, they tend to resist preventive defense measures, such as patching, virus scanning, clearing cache, checking for secure sites before entering credit card information, or paying attention to spear phishing.
    \item \textbf{Ingroup bias}: People are social animals and give preferential treatment to in-group members over out-group ones. 
    An attacker can pretend to be affiliated with  the group to gain trust and influence the decisions of group members. 
    %\item Confirmation bias: People tend to collect information that confirm their preconceptions and discredit information that contradict the initial opinion. Attacks can use confirmation bias to  ‘see what they want to see’ 
\end{itemize}
In marketing and persuasion, \citet{cialdini2007influence} deduced six principles of influence from experimental and field studies based on exploitable personal traits. 
Attackers can also take advantage of the following traits to complete the compromise. 
\begin{itemize}
    \item \textbf{Reciprocity}: 
    \acrshort{se}  attackers frequently offer victims something to set the stage for reciprocity. For example, people are less likely to refuse an inappropriate request from someone who has provided them with a gift in advance. 
    \item \textbf{Social proof}: 
    Individuals are easily influenced by the decisions of a large group of people. An attacker can impersonate the involvement of a victim's friends in order to compel victims to act. 
    \item \textbf{Authority}: 
    People tend to conform to authority, and attackers can exploit that by pretending to be a system administrator or a security expert. 
   \item \textbf{Liking}: 
  Since it is much easier to influence a person who likes you, attackers can attempt to be likeable by being polite and using concepts, languages, and appearances familiar to the target. 
   \item \textbf{Scarcity}: 
   Something in short supply can increase its sense of value. Social engineers may use scarcity (e.g., a limited-time offer) to create a feeling of urgency and spur the victim's action. 
   \item \textbf{Commitments and consistency}: 
   Once people make a choice, they  receive pressure from others and themselves to adhere to it. 
   \acrshort{se}  attacks can induce the victim to make a seemingly insignificant commitment and then escalate the requests. People tend to accept the escalation of commitment as long as the subsequent requests are consistent with the prior commitment. 
\end{itemize}

\subsection{Cognitive Attack}
\label{subsec:Cognitive Attack}

In Section \ref{subsec:Cognitive Vulnerability}, we discuss four types of exploitable vulnerabilities in the cognitive process. 
Cognitive attacks refer to the adversarial processes that involve the exploitation of these human vulnerabilities, as defined in Definition \ref{def:Cognitive Attacks}.   
\begin{definition}[Cognitive Attacks]
\label{def:Cognitive Attacks}
Cognitive attacks are a class of cyber-physical-human processes that manipulate the behaviors of human actors for malicious purposes, including theft or damage of confidential data and disruption or misdirection of the \acrshort{hcps} services, by exploiting their cognitive vulnerabilities. 
\end{definition}
Analogous to cyber kill chains\footnote{
%The cyber kill chain was created by Lockheed Martin in 2011 and explains the steps of numerous typical cyberattacks, as well as the points where the information security team may prevent, detect, or intercept attackers. 
An example cyber kill chains developed by Lockheed Martin in $2011$ can be found at \url{https://www.lockheedmartin.com/en-us/capabilities/cyber/cyber-kill-chain.html}}, we present the kill chain of cognitive attacks in Section \ref{subsubsec:Kill Chain}. 
Then, in Section \ref{subsubsec:Examples Paths}, we zoom into the execution phase of the kill chain and use \acrshort{se}  and \acrshort{idos} attacks as two examples to illustrate the cross-layer attack paths of cognitive attacks. 

\subsubsection{Kill Chain of Cognitive Attacks}
\label{subsubsec:Kill Chain}

\begin{figure}[tbh]
\sidecaption[t]
\includegraphics[width=1 \textwidth]{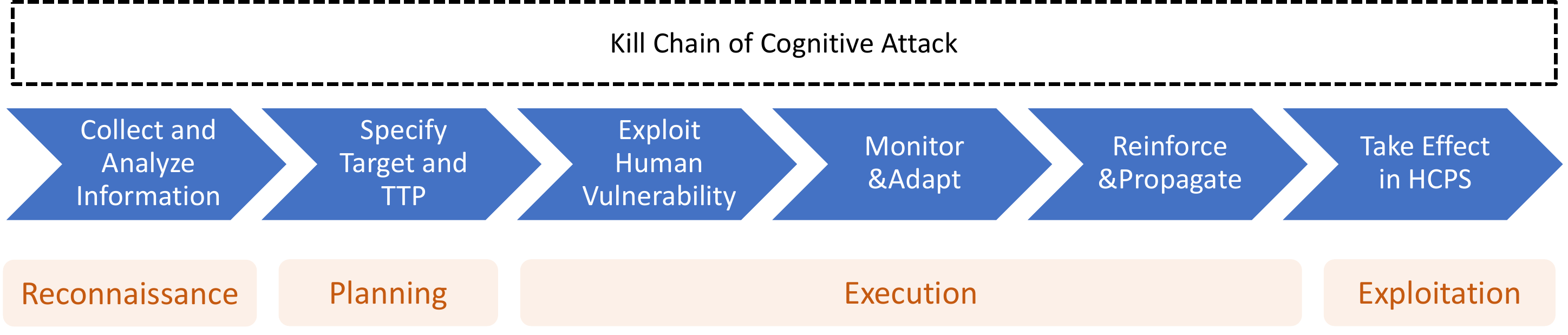}
\caption{
The illustration of the kill chain of cognitive attacks concerning the four-stage attack phases of reconnaissance, planning, execution, and exploitation. 
In the execution stage, cognitive attacks exploit human vulnerabilities, monitor their responses, and adapt the exploitation \acrshort{ttp} accordingly. 
Then, they choose the best attack setting to reinforce the compromise effect and spread it to a group of victims from the initial victim. 
}
\label{fig:killchain}      
\end{figure}

Fig. \ref{fig:killchain} illustrates the conceptual kill chain of cognitive attacks that consists of six stages in blue arrows.  
We map the six kill-chain stages into the four attack phases of the attack stack in Fig. \ref{fig:4stack} at the bottom of Fig. \ref{fig:killchain} in orange.  

In the reconnaissance phase, the attacker analyzes the information collected from the \acrshort{hcps}. 
For example, an attacker can formulate the behavioral baseline of a user or an operator during the reconnaissance. Such a baseline can help the attack identify the human vulnerabilities of the target users and exploit them in the execution phase.  
%; e.g., the attacker may deduce the pattern and determine private email addresses based on public email addresses. 
In the planning phase, the attackers identify the valuable assets and tailor their \acrfull{ttp} accordingly. 
We divide the execution phase into three stages. First, attackers exploit human vulnerabilities directly or indirectly, where we show several examples of exploitation paths in Section \ref{subsubsec:Examples Paths}. 
Second, attackers monitor the human target's responses (e.g., emotions and behaviors) and adapt their \acrshort{ttp}. 
For example, attackers can choose to increase or decrease the attack frequency for cautious and careless users, respectively. 
After the attackers obtain the optimal attack setting, they continue to reinforce the cognitive exploitation and propagate the compromise to other victims related to the initial human target. 
For example, after gaining trust, attackers can ask the initial victim to forward emails to his colleague, who will become less doubtful about the legitimacy of the phishing emails.  
In the exploitation phase, cognitive exploitation begins to take effect in human behaviors and subsequently the cyber and physical layers.

\subsubsection{Examples Paths of Cognitive Attacks}
\label{subsubsec:Examples Paths}

\begin{figure}[tbh]
\sidecaption[t]
\includegraphics[width=1 \textwidth]{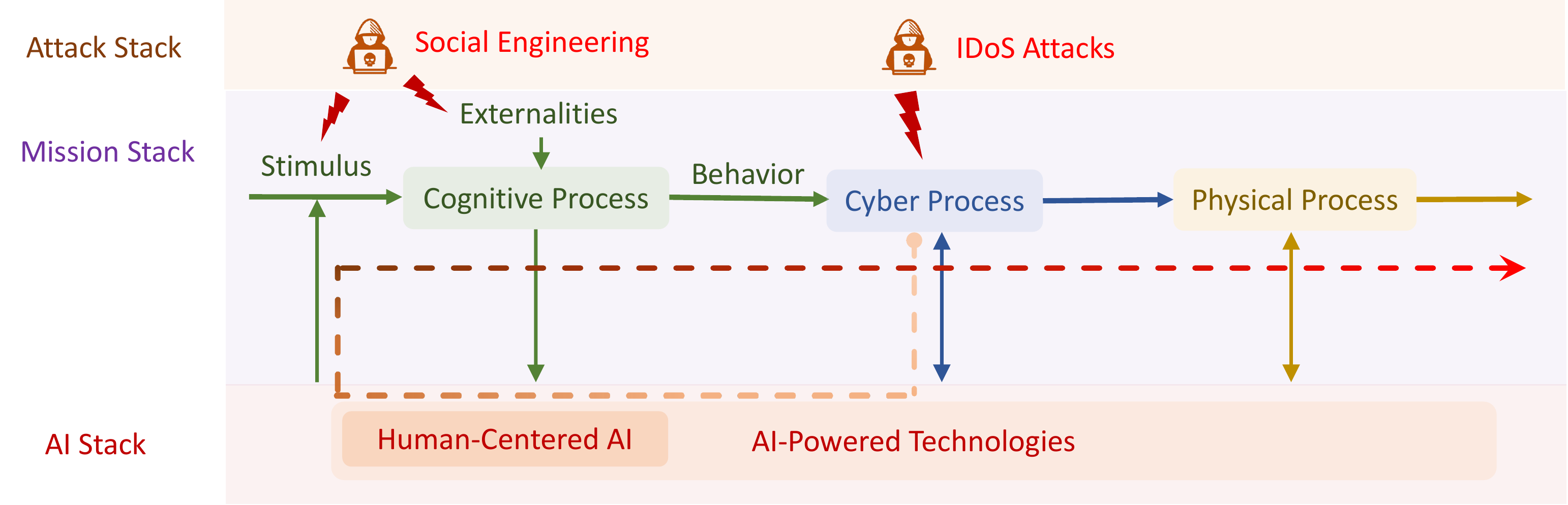}
\caption{
Example paths of cognitive attacks are depicted in the dashed arrow. 
\acrshort{se}  attacks directly exploit human cognitive vulnerabilities by changing stimuli and influencing external factors. 
\acrshort{idos} attacks first compromise the cyber process and use technical compromise as a stepping stone to affect the \acrshort{hmi} (e.g., by creating feints to increase the operator's cognitive load), which indirectly manipulates the cognitive process. 
Both types of cognitive attacks have an impact on the cyber and physical processes. 
}
\label{fig:Cognitiveattack}    
\end{figure}

Exploiting human vulnerabilities is a critical step in the execution attack stage. It can take different forms and involve one or many of the cognitive vulnerabilities described in Section \ref{subsec:Cognitive Vulnerability}. 
In Fig. \ref{fig:Cognitiveattack}, we use \acrshort{se}  and \acrshort{idos} attacks as two examples to illustrate the various forms and procedures of cognitive attacks. 

\acrshort{se}  attacks directly manipulate the stimulus (e.g., phishing content) and external factors (e.g., peer pressure) to compromise human users. 
In this type of cognitive attack, cognitive compromise is used as the stepping stone to enter the \acrshort{cps} and perpetrate the technical compromise. 

However, in other cognitive attacks, the technical compromise may also serve as a precondition to exploit human cognitive vulnerabilities, as illustrated by the attack path of \acrshort{idos} attacks. 
\acrshort{idos} attacks first generate a lot of feint attacks to trigger cyber alerts that are displayed through a \acrshort{hmi}. 
Human operators investigating these alerts in real-time can suffer from \textit{cognitive overload}, which leads to reduced accuracy and speed in processing the alerts. 
Since human operators may be unable to respond to alerts associated with real and significant attacks, these attacks have the potential to disrupt both cyber and physical processes.
Examples of real-world \acrshort{idos} attacks widely exist but are usually implicit in many attack incidents. 
The following three incidents in Examples \ref{example:Sony 2011}, \ref{example:BIPS 2013}, and \ref{example:tesla 2021} use \acrfull{ddos} attacks as a ``smoke shell'' to attract security analysts' attention while simultaneously launching other stealthy attacks.

\begin{svgraybox}
\begin{Example}[Sony PSN Data Breach 2011]
\label{example:Sony 2011}
In April 2011, the \acrfull{soc} of Sony was occupied dealing with a \acrshort{ddos} attack, and they overlooked the extraction of confidential information, including the names, addresses, dates of birth, and passwords, of 77 million PlayStation Network (PSN) customers. 
Authorities in the U.K.  fined the Sony company the equivalent of almost \$400,000 over the failures in the PSN data breach \cite{SonyPSNDDoS}. 
\qed{}
\end{Example}
%\end{svgraybox}

%\begin{svgraybox}
\begin{Example}[BIPS BTC Stolen 2013]
\label{example:BIPS 2013}
Europe’s primary bitcoin payment processor for merchants and free online wallet service, BIPS, became the target of a massive \acrshort{ddos} attack on November 15th, 2013. 
While the \acrshort{soc} was busy to get the system back online, the attacker launched a subsequent attack to gain access and compromise several wallets, resulting in the theft of 1,295 BTC (approximately $1$ million dollars) \cite{BIPSddos}. 
\qed{}
\end{Example}
%\end{svgraybox}

%\begin{svgraybox}
\begin{Example}[Tesla Ransomware 2021]
\label{example:tesla 2021}
A recent thwarted cybersecurity attack against Tesla in 2021  planned to use \acrshort{ddos} attack to divert security analysts' attention from malware that extracted confidential data \cite{idosexam}. 
%bribe a Tesla's employee to transmit malware for data extraction. 
\qed{}
\end{Example}
\end{svgraybox}

\subsection{Cognitive Security}
\label{subsec:Cognitive Security}

\begin{figure}[tbh]
\sidecaption[t]
\includegraphics[width=1 \textwidth]{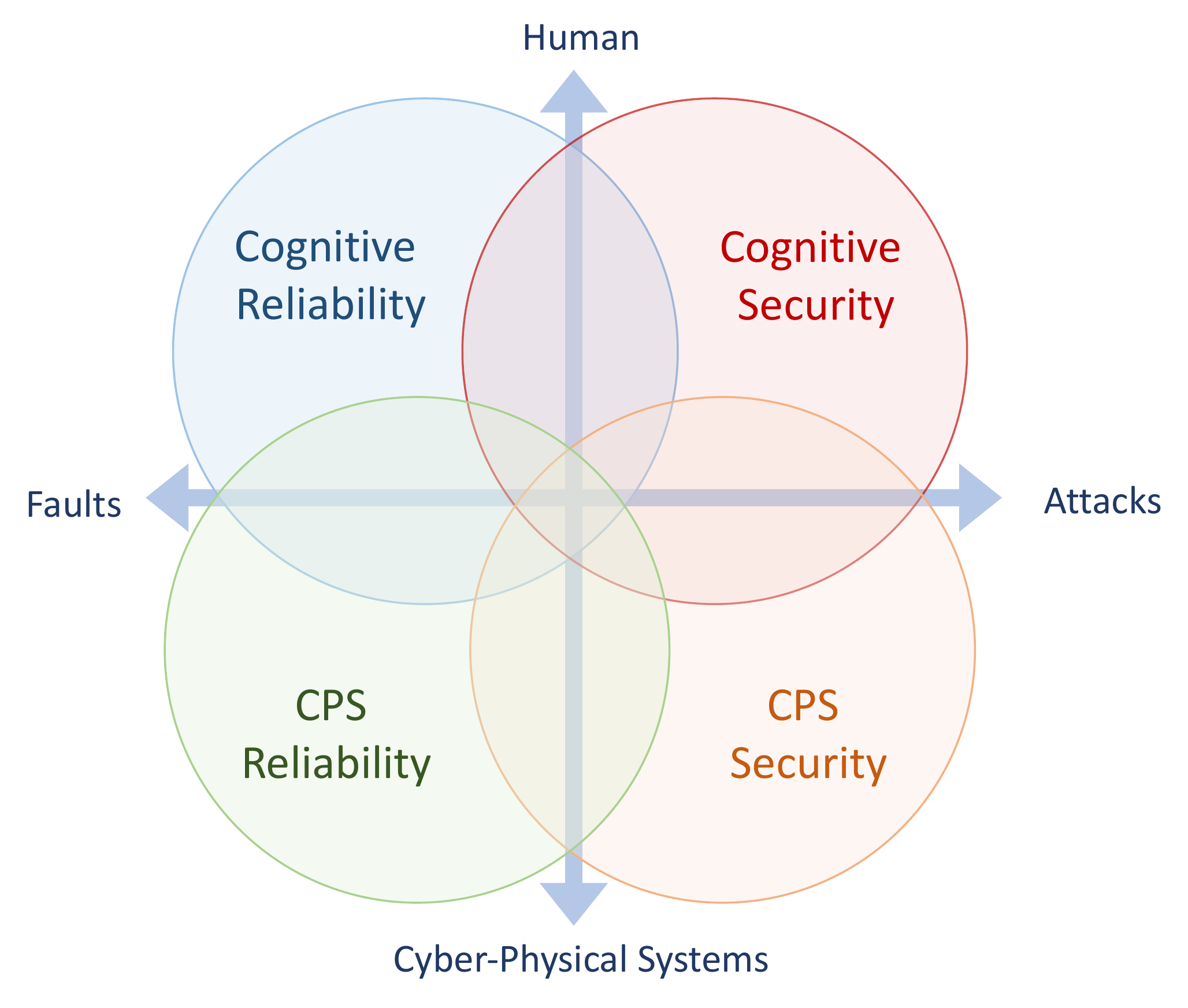}
\caption{
Four quadrants of security and reliability in \acrshortpl{cps} and \acrshortpl{hcps}. 
The horizontal arrows distinguish security and reliability based on whether attacks are present or absent, respectively. 
The vertical arrows distinguish security (or reliability) in \acrshortpl{hcps} or \acrshortpl{cps} based on whether cognitive vulnerabilities are involved or not. 
Cognitive security generalizes \acrshort{se}  and its mitigation through the additional investigation of the cross-layer impact on \acrshortpl{hcps} and the development of human-technical defense technologies to deter the kill chain of cognitive attacks. 
}
\label{fig:Quad}    
\end{figure}

%no figure cited, we talk about cps only. 
The development of the  \acrshort{ai}-powered \acrshort{cps} begins with reliability assurance to guarantee that the system and  \acrshort{ai} technologies can remain trustworthy and efficient under uncertainty, disturbances, and failures. 
The presence of attacks requires us to incorporate defenses, which leads to the concept of security. 
Analogous to reliability and security in \acrshort{cps}, we define and distinguish cognitive reliability and cognitive security in Definitions \ref{def:Cognitive Reliability} and \ref{def:Cognitive Security}, respectively, to form the four quadrants in Fig. \ref{fig:Quad}. 

\begin{definition}[Cognitive Reliability]
\label{def:Cognitive Reliability}
Cognitive reliability is the capacity of a \acrshort{hcps} to maintain the continuity of operations, fulfil a designated mission, and provide services at the desired level under stated conditions, including challenging environments of uncertainty, disturbance, and error. 
%consistently well under uncertainty, disturbance, and failures. 
%without failure
\end{definition}

\begin{definition}[Cognitive Security]
\label{def:Cognitive Security}
Cognitive security is the practice of deploying people, policies, processes, and technologies to withstand cognitive attacks in Definition \ref{def:Cognitive Attacks} and defend essential \acrshort{hcps} components, including humans, critical system structures, services, and sensitive information. 
%protect essential \acrshort{hcps} components (e.g., humans, critical system structures, services, and sensitive information) from cognitive attacks in Definition \ref{def:Cognitive Attacks}. 
\end{definition}

\subsubsection{Cognitive Reliability vs. Cognitive Security}
\label{subsubsec:Cognitive Reliability vs. Cognitive Security}
From Definition \ref{def:Cognitive Reliability}, cognitive reliability focuses on \acrfull{hra} and human-centered system design. 
On the one hand, \acrshort{hra} identifies potential human error events, evaluates the contributing factors, estimates the probability of those errors, and analyzes their impact on the \acrshort{cps}. 
A variety of methods exist for \acrshort{hra}, including techniques based on probabilistic risk analysis and cognition control, where \citet{di2013overview} refers to these two classes of techniques as the first and second generations of \acrshort{hra}, respectively. 
These \acrshort{hra} techniques have been widely used in life-critical industrial control systems (e.g., nuclear power plants) to minimize the adverse consequences of human behaviors on the technical systems.  

On the other hand, human-centered system design includes ergonomics and behavioral science to understand human cognition processes, adapt the system to human factors to enhance usability, and reduce erroneous behaviors. 
For example, based on the behavioral science findings that recognition is significantly easier than recall, text-based challenge-response mechanisms have been applied in user-to-computer authentication as an improvement over unaided password recall \cite{zviran1990cognitive}. 
Other cognitive science results have been used in user authentication and email encryption for usable security \cite{payne2008brief}; e.g., a graphical password emerges because people memorize pictures better than texts. 

%\subsubsection{Cognitive Security}
Cognitive security in Definition \ref{def:Cognitive Security} is a concept associated with cognitive attacks in Section \ref{subsec:Cognitive Attack}. 
Unlike human-induced failures in cognitive reliability, cognitive attacks can directly take advantage of cognitive weaknesses and use them as stepping stones to maximize the attack gain. 
The term ``at the desired level under stated conditions'' in Definition \ref{def:Cognitive Reliability} indicates that cognitive reliability has a specific goal of system efficiency and usability for a set of defined conditions. 
Performance under these conditions enables cognitive functions to ``withstand attacks" or errors up to a certain capacity. 
For attacks that are beyond such capacity, defense methods are needed to provide further protection, as suggested in Definition \ref{def:Cognitive Security}. 
%This capacity is the outcome of the tradeoff between performance and fault tolerance. 
Protection methods come at the cost of reduced efficiency (e.g., false alarms of an \acrfull{ids} disrupting normal operation) and usability (e.g., additional effort needed to comply with security procedures). 
%, we use ``withstand attacks'' in Definition \ref{def:Cognitive Security}. 
% Due to the negative effects of defense, security cannot be a simple add-on but rather a holistic design that quantifies the
Security designs that aim to augment the capacity needs to takes into account the tradeoff among security, efficiency, and usability in a holistic manner. Such a rationale leads to the system-scientific perspectives, which will be discussed in Section \ref{sec:System-Scientific Perspectives for Cognitive Security}. 

\subsubsection{CIA Triad of Cognitive Security}
\label{subsubsec:CIAtriad}
In the context of cognitive security, we discuss confidentiality, integrity, and availability, i.e., the CIA triad, to provide a guide for theoretical foundations.
Confidentiality prevents unauthorized users from accessing the sequential information flow of the cognitive process shown in Fig. \ref{fig:CogProZoomin}. 
Phishing attacks, for example, violate confidentiality by masquerading as legitimate entities and collecting confidential information. 
The attacker can use the information to acquire access credentials, steal data, and deploy malware (e.g., ransomware). 

The integrity of cognitive security assesses whether the cognitive process depicted in Fig. \ref{fig:CogProZoomin}  has been manipulated to induce biased actions. 
Examples of cognitive attacks that compromise integrity include misinformation propagation and belief manipulation techniques (e.g., gaslighting). 

The availability of cognitive security guarantees the normal functioning of the cognitive process in Fig. \ref{fig:CogProZoomin}. 
\acrshort{idos} attacks compromise availability by depleting the limited cognitive resources, including attention, memory, and decision capacity.

\subsubsection{Cognitive and Technical Defenses for Cognitive Security}
\label{subsubsec:Cognitive and Technical Defenses}
\begin{figure}[tbh]
\sidecaption[t]
\includegraphics[width=1 \textwidth]{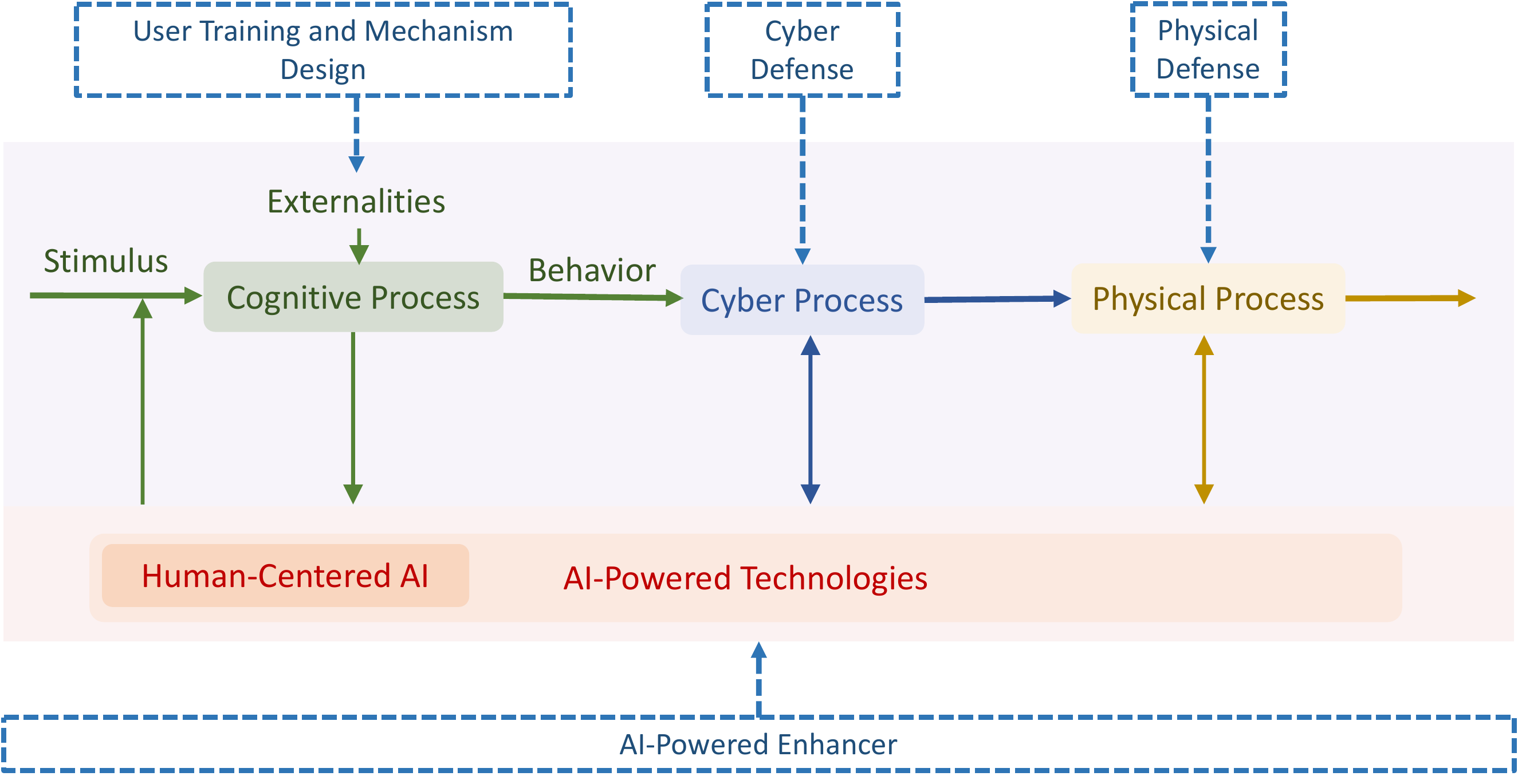}
\caption{
The defense methods in blue dashed boxes to break the kill chain of cognitive attacks and achieve cognitive security. 
The cyber and physical defenses take effect on the cyber and physical processes, respectively. 
User training and human-centered mechanism design affect the cognitive process through externalities. Incorporating the  \acrshort{ai} stack, cognitive security includes the  \acrshort{ai}-powered enhancer to protect \acrshortpl{hcps} from the adversarial exploitation of  \acrshort{ai} technologies. 
}
\label{fig:cognitivesecurity}       % Give a unique label
\end{figure}

Cognitive security synthesizes cognitive and technical defenses to break the kill chain of cognitive attacks, as shown in Fig. \ref{fig:cognitivesecurity}.  
Since cognitive attacks include technical exploitation across the cyber and physical layers, as shown in the IDoS attack path in Section \ref{subsubsec:Examples Paths}, cyber and physical defenses are indispensable components.  

Besides the cyber and physical defenses, cognitive security also focuses on defending humans and  \acrshort{ai} technologies. 
On the one hand, user training can help users gain security knowledge, raise security awareness, and cultivate critical thinking, which consolidates the cognitive process. 
On the other hand, we can design mechanisms to avoid selfish behaviors and incentivize users to behave securely.

 \acrshort{ai}-technologies in  \acrshort{ai} stack can also be susceptible to cognitive attacks. %, as shown by the IDoS attack path in Section \ref{subsubsec:Examples Paths}. 
Keeping the security requirement in mind, we need to design an  \acrshort{ai}-powered enhancer to guarantee the security and resilience of the  \acrshort{ai}-powered technologies. 
In particular, human-centered  \acrshort{ai} technologies in  \acrshort{ai} stack  need to not only tilt the cognitive biases when augmenting human capacities, but also be aware of and defend against the cognitive attacks. 
For example,  \acrshort{ai}-enabled security assistive technologies can argument \acrshort{hmi} to defend against the threat from \acrshort{idos} attacks. 
To achieve the goal, we fuse various biosensor data, including optical, motion, and acoustic signals and electrical biosignals, to reflect the status of the current cognitive process in real-time. 
These signals can be fed back to the  \acrshort{ai}-enhancer for adaptive control.

We can analogize the human process augmented by  \acrshort{ai}-powered technologies in Fig. \ref{fig:cognitivesecurity} as children under the supervision of parents. 
While user training and mechanism design teach and motivate children to behave securely, these methods are subject to social and cognitive limitations. 
To break the limitation, the \acrshort{ai}-powered enhancer alternatively lets parents be aware of and protect against the potential attacks on their children.

\subsubsection{Scope of Cognitive Security} %Related Concepts and 
\label{subsubsec:Scope of Cognitive Security}

To better illustrate the scope of cognitive security, we distinguish the following related concepts from cognitive security in Definition \ref{def:Cognitive Security} through two running examples. 

First, the exploitation and defense of human cognition are essential components of cognitive security, which is significantly different from the realm of CPS security and creates a new battlefield of \acrshort{hcps} security. 
%For example, for insider threat, previous works focus on t

Second, as illustrated in Fig. \ref{fig:Quad}, cognitive security generalizes the concept of \acrshort{se} and cognitive hacking \cite{cybenko2002cognitive} that usually rely on weaponized information and exploit vulnerabilities of mental operation. 
On top of the cognitive manipulation in the human-human interaction, we further investigate its impacts and dependency on the \acrshort{cps} during the human-machine interaction. 
As illustrated in the \acrshort{idos} example, technical exploitation can exacerbate the attentional vulnerability. Such human-technical attacks that directly exploit human cognition limitations have not been considered in \acrshort{se} and cognitive hacking. 

Third, cognitive security in this book does not aim to enhance humans at the neuroscience or biological level through the means of medications or medical therapies. 
Instead, we focus on designing mechanisms and assistive technologies to non-intrusively influence and externally affect human cognitive processes, change their behaviors, and ultimately enhance the security of the entire \acrshort{hcps}. 

Fourth, cognitive security does not aim to answer human-engineering questions such as how to craft convincing misinformation or how to design an anti-phishing interface. 
Instead, we leverage theories and empirical results from behavioral science and psychology at the system-level. 
%For example, as we will demonstrate  in Chapter \ref{chap: ADVERT}, we develop a data-driven approach to selecting  the optimal anti-phishing interface, given the assumption that there is a pre-established library of interfaces. 
%from a pre-established library of interfaces that are potentially constructed using human engineering. 

Fifth, technical solutions for cognitive security are aware of and adapt to adverse exploitation of cognitive vulnerabilities. 
For example, under the threat of \acrshort{idos} attacks, it may be insufficient to adopt the traditional method of alert triage to indirectly reduce cognitive load. 
Therefore, we further de-emphasize alerts strategically to directly manage attention and make the alert inspection process compatible with each operator's cognitive load. 

Sixth, cognitive defense is an essential but not the only way to achieve cognitive security. 
We may not need to defend against all types of cognitive attacks. It is sufficient to break the kill chain of cognitive attacks. 
Moreover, since some types of cognitive attacks are not defensible or too costly to defend, we can focus on compensation and correction mechanisms to reduce the risk and impact of the attacks. 

Finally, cognitive security in this book does not mean developing security measures to have human cognitive capacities (e.g., cognitive computing \acrshort{soc} \cite{demertzis2018next}, phishing detector \cite{garces2019detection}, and incident management systems \cite{andrade2019cognitive}), nor does it mean the exploitation of human cognitive strengths (e.g.,  creativity and flexibility) to augment technical defense systems. 
Moreover, cognitive security does not focus on increasing the usability of the security measures \cite{greenstadt2008cognitive,payne2008brief}, despite its contribution to cognitive security.

\section{System-Scientific Perspectives for Cognitive Security}
\label{sec:System-Scientific Perspectives for Cognitive Security}

Following Sections \ref{sec: AI-Powered Human-Cyber-Physical Systems} and \ref{sec:Cognitive Security in HCPS}, humans are an essential component in accomplishing \acrshort{cps} missions, and cognitive security needs to incorporate rather than eschew human cognition. 
Human cognition, however, is of high complexity, uncertainty, and individual difference, which leads to the following research questions.  
How to incorporate established theories of human factors into \acrshortpl{hcps} for defensive purposes? 
How to customize the defense and make it adaptive to various security applications? 
What are useful metrics and measures to quantify the impact of human factors on a \acrshort{cps} and the effectiveness of the defense methods? 

This book adopts system-scientific perspectives, including decision theory, optimization, game theory,  \acrshort{ai}, physiological metrics, and psychology theories to address these questions for the following reasons. 
\begin{itemize}
    \item First, as illustrated in Section \ref{subsec:Cognitive Attack}, cognitive attacks occur at the \textit{cognition level} (e.g., \acrshort{se}) and the system level (e.g., \acrshort{idos} attacks) rather than  the \textit{neuronal level} through medicine. Thus, the attack and defense interactions need to be investigated at the \textit{system level} to understand the entire attack phase, anticipate the interaction of attacks with the \acrshort{hcps}, and find defensible points among these phases. 
    \item  Second, cognitive attacks (e.g., \acrshort{idos} attacks) exploit distinct vulnerabilities in human, cyber, and physical processes. 
    %Different science is needed for different components. 
    A single tool is not sufficient to deter, detect, and prevent these cross-layer, cross-disciplinary attacks. 
    To safeguard  diverse vulnerabilities, we need interdisciplinary tools to synthesize defense methods from multiple areas. 
    \item Third, since cognitive attacks have an impact on human, cyber, and physical processes, it is important to quantify the risk to each component, the prorogation of the risk, and the risk to the entire \acrshort{hcps} \cite{huang2017large,huang2018distributed,chen2019game}. 
    Moreover, we need to design optimal controllables to defend \acrshortpl{hcps} cost-effectively. 
    Therefore, we need scientific and quantitative approaches to quantify security metrics and bounds, assess tradeoffs, characterize fundamental limits, and design optimal control \cite{wiener2019cybernetics,meadows2008thinking,xu2016cross}. 
    \item Fourth, since both cognitive attackers and defenders are intelligent players, they take actions based on the predictions of the others. Game theory becomes a natural tool to capture the interactions in adversarial environment for quantitative analysis and design \cite{manshaei2013game,zhu2015game,huang2020dynamic,zhu2018game}. 
    \item Fifth, models may not accurately describe \acrshort{hcps} components (e.g., human cognition) that are highly dynamic and sometimes unobservable. Data collected by biosensors (e.g., electroencephalogram (EEG) and eye-tracking devices) and  \acrshort{ai} can be incorporated to provide system-level adaptive solutions in response to the measurable observations. 
    \item Finally, theories from psychology and social science can be adopted to understand perception, attention, memory, and decisions. As a multi-disciplinary methods, system-scientific approach incorporate those findings and results to create  holistic hybrid data-driven and model-based system frameworks of \acrshortpl{hcps}. 
\end{itemize}

\subsection{Advantages of System-Scientific Approaches}
\label{sec:Benefits of System-Scientific Approaches}

System-scientific approaches provide a new paradigm for cognitive security and bring the following advantages. 
%combine the benefits of system theory and scientific methods. 

%The first three are more system

\begin{itemize}
    \item \textbf{Emergence}: 
    Emergence widely appears in philosophy, psychology, and art, where ``the whole is something else than the sum of its parts'' \cite{koffka2013principles}\footnote{often misquoted as `the whole is greater than the sum of its parts'.}.  
    As shown in Fig. \ref{fig:TwoTrends}, the close integration of  \acrshort{ai} and humans in \acrshortpl{cps} leads to the new concept of \acrshortpl{hcps}. The interactions among human, cyber, physical, and  \acrshort{ai} components create new attack surfaces and attack paths, which potentially amplifies the human cognitive vulnerabilities and promotes the need for research in the emerging field of cognitive security. 
    
    \item \textbf{Black Box and Function Simulation}: A system perspective focuses on the input and output (or the transfer characteristics) of the system, where the system itself is treated as a black box without any knowledge of its internal workings. 
    By treating human cognition systems as black boxes, we focus on the \textit{behavior-level impact} rather than the \textit{cognitive- or neuro-level mechanisms} that can be complicated and not well-understood. 
    The input-output view also enables us to simulate a system's function without establishing the internal model of the system. 
    
    \item \textbf{Modular and Multi-Scale Design}: 
    A system can be divided into different levels of subsystems that compose a multi-scale system model. 
    For example, the \acrshort{hcps} system consists of human, cyber, and physical subsystems, where the human subsystem further contains cognition systems consisting of perception, attention, memory, and mental operations. 
    We can zoom into the proper level of subsystems based on our goal. 
    Since these subsystems interact through their inputs and outputs, each subsystem can be replaced and revised to achieve a modular design. 
    
%     \item \textbf{Coordination and Tradeoff}: 
%   The system-level modeling creates a holistic view of its interacting components, constraints, and trade-offs among multiple objectives. 
%     {\bf I DO NOT QUITE UNDERSTAND THIS PARAGRAPH: For example, emphasizing technical advances may make them complicated and challenging for operators to understand, which leads to worse performance in the end. }
    
    \item \textbf{Quantitative Strategies and Optimization}: 
     The system-scientific models enable a quantitative description of the situation  and thus the formulation of  optimal design problems that lead to cost-effective security mechanisms. The integration of multiple system-scientific approaches, including deterministic and stochastic methods, data-driven and model-based tools, and static and dynamic frameworks, strengthens their strong points and mitigates their weak points. 

\end{itemize}

 \bibliographystyle{author/spbasic}
% %\bibliographystyle{spphys}
% %\bibliographystyle{spmpsci}
 \bibliography{CogSec}

\begin{thebibliography}{57}
\providecommand{\natexlab}[1]{#1}
\providecommand{\url}[1]{{#1}}
\providecommand{\urlprefix}{URL }
\expandafter\ifx\csname urlstyle\endcsname\relax
  \providecommand{\doi}[1]{DOI~\discretionary{}{}{}#1}\else
  \providecommand{\doi}{DOI~\discretionary{}{}{}\begingroup
  \urlstyle{rm}\Url}\fi
\providecommand{\eprint}[2][]{\url{#2}}

\bibitem[{Andrade et~al.(2019)Andrade, Torres, and
  Cadena}]{andrade2019cognitive}
Andrade R, Torres J, Cadena S (2019) Cognitive security for incident management
  process. In: International Conference on Information Technology \& Systems,
  Springer, pp 612--621

\bibitem[{Arons(1992)}]{arons1992review}
Arons B (1992) A review of the cocktail party effect. Journal of the American
  Voice I/O Society 12(7):35--50

\bibitem[{Barmer et~al.(2021)Barmer, Dzombak, Gaston, Palat, Redner, Smith, and
  Wohlbier}]{Barmer2021}
Barmer H, Dzombak R, Gaston M, Palat V, Redner F, Smith T, Wohlbier J (2021)
  {Scalable AI} \doi{10.1184/R1/16560273.v1},
  \urlprefix\url{https://kilthub.cmu.edu/articles/report/Scalable_AI/16560273}

\bibitem[{Biddle et~al.(2012)Biddle, Chiasson, and
  Van~Oorschot}]{biddle2012graphical}
Biddle R, Chiasson S, Van~Oorschot PC (2012) Graphical passwords: Learning from
  the first twelve years. ACM Computing Surveys (CSUR) 44(4):1--41

\bibitem[{Bruce et~al.(2003)Bruce, Green, and Georgeson}]{bruce2003visual}
Bruce V, Green PR, Georgeson MA (2003) Visual perception: Physiology,
  psychology, \& ecology. Psychology Press

\bibitem[{Chaczko et~al.(2020)Chaczko, Kulbacki, Gudzbeler, Alsawwaf,
  Thai-Chyzhykau, and Wajs-Chaczko}]{chaczko2020exploration}
Chaczko Z, Kulbacki M, Gudzbeler G, Alsawwaf M, Thai-Chyzhykau I, Wajs-Chaczko
  P (2020) Exploration of explainable ai in context of human-machine interface
  for the assistive driving system. In: Asian Conference on Intelligent
  Information and Database Systems, Springer, pp 507--516

\bibitem[{Chen and Zhu(2019)}]{chen2019game}
Chen J, Zhu Q (2019) A game-and decision-theoretic approach to resilient
  interdependent network analysis and design. Springer

\bibitem[{Cialdini(2007)}]{cialdini2007influence}
Cialdini RB (2007) Influence: The psychology of persuasion, vol~55. Collins New
  York

\bibitem[{Commission et~al.(2021)Commission, for Research, Innovation, Breque,
  De~Nul, and Petridis}]{industry5}
Commission E, for Research DG, Innovation, Breque M, De~Nul L, Petridis A
  (2021) Industry 5.0 : towards a sustainable, human-centric and resilient
  European industry. Publications Office, \doi{doi/10.2777/308407}

\bibitem[{Cox et~al.(2020)Cox, Zhu, and Balcetis}]{cox2020stuck}
Cox EB, Zhu Q, Balcetis E (2020) Stuck on a phishing lure: Differential use of
  base rates in self and social judgments of susceptibility to cyber risk.
  Comprehensive Results in Social Psychology 4(1):25--52

\bibitem[{Cybenko et~al.(2002)Cybenko, Giani, and
  Thompson}]{cybenko2002cognitive}
Cybenko G, Giani A, Thompson P (2002) Cognitive hacking: A battle for the mind.
  Computer 35(8):50--56

\bibitem[{Deese(1959)}]{deese1959prediction}
Deese J (1959) On the prediction of occurrence of particular verbal intrusions
  in immediate recall. Journal of experimental psychology 58(1):17

\bibitem[{Demertzis et~al.(2018)Demertzis, Kikiras, Tziritas, Sanchez, and
  Iliadis}]{demertzis2018next}
Demertzis K, Kikiras P, Tziritas N, Sanchez SL, Iliadis L (2018) The next
  generation cognitive security operations center: network flow forensics using
  cybersecurity intelligence. Big Data and Cognitive Computing 2(4):35

\bibitem[{Di~Pasquale et~al.(2013)Di~Pasquale, Iannone, Miranda, and
  Riemma}]{di2013overview}
Di~Pasquale V, Iannone R, Miranda S, Riemma S (2013) An overview of human
  reliability analysis techniques in manufacturing operations. Operations
  management pp 221--240

\bibitem[{Doghri et~al.(2022)Doghri, Saddoud, and
  Chaari~Fourati}]{doghri2022cyber}
Doghri W, Saddoud A, Chaari~Fourati L (2022) Cyber-physical systems for
  structural health monitoring: sensing technologies and intelligent computing.
  The Journal of Supercomputing 78(1):766--809

\bibitem[{Elgendi et~al.(2018)Elgendi, Kumar, Barbic, Howard, Abbott, and
  Cichocki}]{elgendi2018subliminal}
Elgendi M, Kumar P, Barbic S, Howard N, Abbott D, Cichocki A (2018) Subliminal
  priming—state of the art and future perspectives. Behavioral Sciences
  8(6):54

\bibitem[{Fisher(2013)}]{SonyPSNDDoS}
Fisher D (2013) Sony fined £250,000 by uk over failures in playstation network
  breach.
  \urlprefix\url{https://threatpost.com/sony-fined-250000-uk-over-failures-playstation-network-breach-012413/77446/}

\bibitem[{Garc{\'e}s et~al.(2019)Garc{\'e}s, Cazares, and
  Andrade}]{garces2019detection}
Garc{\'e}s IO, Cazares MF, Andrade RO (2019) Detection of phishing attacks with
  machine learning techniques in cognitive security architecture. In: 2019
  International Conference on Computational Science and Computational
  Intelligence (CSCI), IEEE, pp 366--370

\bibitem[{Greenstadt and Beal(2008)}]{greenstadt2008cognitive}
Greenstadt R, Beal J (2008) Cognitive security for personal devices. In:
  Proceedings of the 1st ACM workshop on Workshop on AISec, pp 27--30

\bibitem[{Griffor et~al.(2017)Griffor, Greer, Wollman, Burns
  et~al.}]{griffor2017framework}
Griffor ER, Greer C, Wollman DA, Burns MJ, et~al. (2017) Framework for
  cyber-physical systems: Volume 1, overview

\bibitem[{Groshev et~al.(2021)Groshev, Guimar{\~a}es, Mart{\'\i}n-P{\'e}rez,
  and de~la Oliva}]{groshev2021toward}
Groshev M, Guimar{\~a}es C, Mart{\'\i}n-P{\'e}rez J, de~la Oliva A (2021)
  Toward intelligent cyber-physical systems: Digital twin meets artificial
  intelligence. IEEE Communications Magazine 59(8):14--20

\bibitem[{Hamann(2001)}]{hamann2001cognitive}
Hamann S (2001) Cognitive and neural mechanisms of emotional memory. Trends in
  cognitive sciences 5(9):394--400

\bibitem[{Huang and Zhu(2022)}]{huang2022radams}
Huang L, Zhu Q (2022) Radams: Resilient and adaptive alert and attention
  management strategy against informational denial-of-service (idos) attacks.
  Computers \& Security 121:102844

\bibitem[{Huang et~al.(2017)Huang, Chen, and Zhu}]{huang2017large}
Huang L, Chen J, Zhu Q (2017) A large-scale markov game approach to dynamic
  protection of interdependent infrastructure networks. In: International
  Conference on Decision and Game Theory for Security, Springer, pp 357--376

\bibitem[{Huang et~al.(2018)Huang, Chen, and Zhu}]{huang2018distributed}
Huang L, Chen J, Zhu Q (2018) Distributed and optimal resilient planning of
  large-scale interdependent critical infrastructures. In: 2018 Winter
  Simulation Conference (WSC), IEEE, pp 1096--1107

\bibitem[{Huang et~al.(2022)Huang, Jia, Balcetis, and Zhu}]{huang2022advert}
Huang L, Jia S, Balcetis E, Zhu Q (2022) Advert: An adaptive and data-driven
  attention enhancement mechanism for phishing prevention. IEEE Transactions on
  Information Forensics and Security 17:2585--2597

\bibitem[{Huang et~al.(2020{\natexlab{a}})Huang, Chen, Jin, and
  Lau}]{huang2020detecting}
Huang W, Chen X, Jin R, Lau N (2020{\natexlab{a}}) Detecting cognitive hacking
  in visual inspection with physiological measurements. Applied ergonomics
  84:103022

\bibitem[{Huang et~al.(2020{\natexlab{b}})Huang, Chen, Huang, and
  Zhu}]{huang2020dynamic}
Huang Y, Chen J, Huang L, Zhu Q (2020{\natexlab{b}}) Dynamic games for secure
  and resilient control system design. National Science Review 7(7):1125--1141

\bibitem[{Johnston and Dark(1986)}]{johnston1986selective}
Johnston WA, Dark VJ (1986) Selective attention. Annual review of psychology

\bibitem[{Kaivanto(2014)}]{kaivanto2014effect}
Kaivanto K (2014) The effect of decentralized behavioral decision making on
  system-level risk. Risk Analysis 34(12):2121--2142

\bibitem[{Kanwisher(1987)}]{kanwisher1987repetition}
Kanwisher NG (1987) Repetition blindness: Type recognition without token
  individuation. Cognition 27(2):117--143

\bibitem[{Koffka(2013)}]{koffka2013principles}
Koffka K (2013) Principles of Gestalt psychology. Routledge

\bibitem[{Lindsay(2020)}]{lindsay2020attention}
Lindsay GW (2020) Attention in psychology, neuroscience, and machine learning.
  Frontiers in computational neuroscience 14:29

\bibitem[{Manshaei et~al.(2013)Manshaei, Zhu, Alpcan, Bac{\c{s}}ar, and
  Hubaux}]{manshaei2013game}
Manshaei MH, Zhu Q, Alpcan T, Bac{\c{s}}ar T, Hubaux JP (2013) Game theory
  meets network security and privacy. ACM Computing Surveys (CSUR) 45(3):1--39

\bibitem[{Marot et~al.(2022)Marot, Rozier, Dussartre, Crochepierre, and
  Donnot}]{marottowards}
Marot A, Rozier A, Dussartre M, Crochepierre L, Donnot B (2022) Towards an ai
  assistant for power grid operators pp 79--95

\bibitem[{Meadows(2008)}]{meadows2008thinking}
Meadows DH (2008) Thinking in systems: A primer. chelsea green publishing

\bibitem[{Molden(2014)}]{molden2014understanding}
Molden DC (2014) Understanding priming effects in social psychology. Guilford
  Publications

\bibitem[{Pan and Yang(2009)}]{pan2009survey}
Pan SJ, Yang Q (2009) A survey on transfer learning. IEEE Transactions on
  knowledge and data engineering 22(10):1345--1359

\bibitem[{Payne and Edwards(2008)}]{payne2008brief}
Payne BD, Edwards WK (2008) A brief introduction to usable security. IEEE
  Internet Computing 12(3):13--21

\bibitem[{Radha and Reddy(2012)}]{radha2012survey}
Radha V, Reddy DH (2012) A survey on single sign-on techniques. Procedia
  Technology 4:134--139

\bibitem[{Rohleder(2019)}]{Rohleder_2019}
Rohleder K (2019) Cognitive biases as vulnerabilities.
  \urlprefix\url{https://www.linkedin.com/pulse/cognitive-biases-vulnerabilities-krinken-rohleder/}

\bibitem[{Salau et~al.(2022)Salau, Rawal, and Rawat}]{salau2022recent}
Salau B, Rawal A, Rawat DB (2022) Recent advances in artificial intelligence
  for wireless internet of things and cyber-physical systems: A comprehensive
  survey. IEEE Internet of Things Journal

\bibitem[{Sasse et~al.(2001)Sasse, Brostoff, and
  Weirich}]{sasse2001transforming}
Sasse MA, Brostoff S, Weirich D (2001) Transforming the ‘weakest link’—a
  human/computer interaction approach to usable and effective security. BT
  technology journal 19(3):122--131

\bibitem[{Sawyer and Hancock(2018)}]{sawyer2018hacking}
Sawyer BD, Hancock PA (2018) Hacking the human: The prevalence paradox in
  cybersecurity. Human factors 60(5):597--609

\bibitem[{Schacter(2002)}]{schacter2002seven}
Schacter DL (2002) The seven sins of memory: How the mind forgets and
  remembers. HMH

\bibitem[{Shapiro et~al.(1997)Shapiro, Raymond, and
  Arnell}]{shapiro1997attentional}
Shapiro KL, Raymond JE, Arnell KM (1997) The attentional blink. Trends in
  cognitive sciences 1(8):291--296

\bibitem[{Simons and Rensink(2005)}]{simons2005change}
Simons DJ, Rensink RA (2005) Change blindness: Past, present, and future.
  Trends in cognitive sciences 9(1):16--20

\bibitem[{Song et~al.(2022)Song, Lyu, Zhang, Wang, Zhang, and Ma}]{Song2022}
Song J, Lyu D, Zhang Z, Wang Z, Zhang T, Ma L (2022) When cyber-physical
  systems meet ai: A benchmark, an evaluation, and a way forward. ICSE 2022
  SEIP

\bibitem[{Southurst(2013)}]{BIPSddos}
Southurst J (2013) Bitcoin payment processor bips attacked, over \$1 million
  stolen.
  \urlprefix\url{https://www.coindesk.com/markets/2013/11/25/bitcoin-payment-processor-bips-attacked-over-1-million-stolen/}

\bibitem[{Team(2021)}]{idosexam}
Team SN (2021) Russian national pleads guilty after trying to hack a human at
  tesla.
  \urlprefix\url{https://www.secureworld.io/industry-news/tesla-hacker-charges-arrested}

\bibitem[{Tipper(1985)}]{tipper1985negative}
Tipper SP (1985) The negative priming effect: Inhibitory priming by ignored
  objects. The quarterly journal of experimental psychology 37(4):571--590

\bibitem[{Vitevitch(2003)}]{vitevitch2003change}
Vitevitch MS (2003) Change deafness: the inability to detect changes between
  two voices. Journal of Experimental Psychology: Human Perception and
  Performance 29(2):333

\bibitem[{Wiener(2019)}]{wiener2019cybernetics}
Wiener N (2019) Cybernetics or Control and Communication in the Animal and the
  Machine. MIT press

\bibitem[{Xu and Zhu(2016)}]{xu2016cross}
Xu Z, Zhu Q (2016) Cross-layer secure cyber-physical control system design for
  networked 3d printers. In: 2016 American Control Conference (ACC), IEEE, pp
  1191--1196

\bibitem[{Zhu and Basar(2015)}]{zhu2015game}
Zhu Q, Basar T (2015) Game-theoretic methods for robustness, security, and
  resilience of cyberphysical control systems: games-in-games principle for
  optimal cross-layer resilient control systems. IEEE Control Systems Magazine
  35(1):46--65

\bibitem[{Zhu and Rass(2018)}]{zhu2018game}
Zhu Q, Rass S (2018) Game theory meets network security: A tutorial. In:
  Proceedings of the 2018 ACM SIGSAC Conference on Computer and Communications
  Security, pp 2163--2165

\bibitem[{Zviran and Haga(1990)}]{zviran1990cognitive}
Zviran M, Haga WJ (1990) Cognitive passwords: The key to easy access control.
  Computers \& Security 9(8):723--736

\end{thebibliography}

%\input{references}
% \include{chapter2}
% \include{chapter3}
% \include{chapter4}
% \include{chapter5}
% \include{chapter6}
% \include{chapter7}

%\include{appendix}

%\backmatter%%%%%%%%%%%%%%%%%%%%%%%%%%%%%%%%%%%%%%%%%%%%%%%%%%%%%%%
%\include{glossary}
%\include{solutions}

%\printindex

%%%%%%%%%%%%%%%%%%%%%%%%%%%%%%%%%%%%%%%%%%%%%%%%%%%%%%%%%%%%%%%%%%%%%%

\end{document}